\documentclass[twocolumn,showpacs,preprintnumbers,amsmath,amssymb]{revtex4}
\usepackage{dcolumn}
\usepackage{amsmath}
\usepackage{graphicx}
\usepackage{latexsym}
\usepackage{amsfonts}
\usepackage{amssymb}
\usepackage{color}
\DeclareGraphicsExtensions{.pdf,.gif,.jpg}
\newcommand{\be}{\begin{equation}}
\newcommand{\ee}{\end{equation}}
\newcommand{\beq}{\begin{eqnarray}}
\newcommand{\eeq}{\end{eqnarray}}

\begin{document}

\newcommand\xmas{\mbox{$x_{\rm mas}$}}
\newcommand\rmas{\mbox{$r_{\rm mas}$}}
\newcommand\Xmas{\mbox{$X_{\rm mas}$}}
\newcommand\Hinf{\mbox{${\rm H}_{\rm inf}$}}
\newcommand\Omegainf{\mbox{$\Omega_{\rm inf}$}}
\newcommand\Omegabg{\mbox{$\Omega_{\rm bg}$}}
\newcommand\Hbg{\mbox{${\rm H}_{\rm bg}$}}
\newcommand\rhobg{\mbox{$\rho_{\rm bg}$}}
\newcommand\rhorbg{\mbox{$\rho^{\rm bg}_{\rm r}$}}
\newcommand\rhoinf{\mbox{$\rho_{\rm inf}$}}
\newcommand\kinf{\mbox{$k_{\rm inf}$}}
\newcommand\kbg{\mbox{$k_{\rm bg}$}}
\newcommand\ainf{\mbox{$a_{\rm inf}$}}
\newcommand\abg{\mbox{$a_{\rm bg}$}}
\newcommand\xpbg{\mbox{$x_{\rm p}^{\rm bg}$}}
\newcommand\xpinf{\mbox{$x_{\rm p}^{\rm inf}$}}
\newcommand\xmasbg{\mbox{$x_{\rm mas}^{\rm bg}$}}
\newcommand\xmasinf{\mbox{$x_{\rm mas}^{\rm inf}$}}
\newcommand\thetainf{\mbox{$\theta^{\rm inf}$}}
\newcommand\thetabg{\mbox{$\theta^{\rm bg}$}}
\newcommand\xHbg{\mbox{$1/{\rm H}_{\rm bg}$}}
\newcommand\xHinf{\mbox{$1/{\rm H}_{\rm inf}$}}

\title{All-electron study of InAs and GaAs wurtzite: structural and electronic properties}

\author{Zeila Zanolli}
\author{Ulf von Barth}
\affiliation{Department of Phisics, Lund University, S\"olvegatan 14 A, 223 62 Lund, Sweden}

\date{\today}

\begin{abstract}

The structural and electronic properties of the wurtzite phase of the InAs and GaAs compounds are, for the first time, studied within the framework of Density Functional Theory (DFT). We used the full-potential linearized augmented plane wave (LAPW) method and the local density approximation (LDA) for exchange and correlation and compared the results to the corresponding pseudopotential calculations.

From the structural optimization of the wurtzite polymorph of InAs we found that the $c/a$ ratio is somewhat greater than the ideal one and that the internal parameter $u/c$ has a value slightly smaller than the ideal one. In the all-electron approach the wurtzite polymorph has a smaller equilibrium volume per InAs pair and a higher binding energy when compared to the zinc-blende phase whereas the situation is reversed in the pseudo treatment. The energy differences are, however, smaller than the accuracy of standard density-functional codes ($\sim 30$~meV) and a theoretical prediction of the relative stability of the two phases cannot be made.   

In order to investigate the possibility of using an LDA calculation as a starting point for many-body calculations of excitations properties, we here also present the band-structures of these materials. The bands are calculated with and without relativistic effects. In InAs we find that the energy gaps of both polymorphs are positive when obtained from a non-relativistic calculation and negative otherwise.

For both semiconductors, we determine the spin-orbit splittings for the zinc-blende and the wurtzite phases as well as the crystal-field splittings for the new wurtzite polymorphs.
\end{abstract}


\maketitle

\section{Introduction}
Nanowires (NW:s) are attracting an increasing amount of attention in the scientific community because they offer the possibility of investigating fundamental physical properties at the nano scale and because of the numerous possible applications in electronic and photonic devices. In particular, InAs-based nanowires have already been used to improve electronics in 1D as, for instance, in the case of resonant tunneling diodes \cite{Biork2002} and single-electron transistors \cite{Thelander2003}. 
These NW:s, typically grown on an (111)B InAs substrate via Chemical Beam Epitaxy (CBE), have a wurtzite crystal structure in contrast to the stable phase at normal pressure and temperature which  is the zinc-blende structure ($zb, 3C$) with the space group $F\overline43m$ ($T^2_d$). 
Despite the successes achieved so far with InAs-based NW:s, it has recently been realized that their wurtzite structure is an important factor in the interpretation of experimental data obtained from measurements of photo-luminescence (PL) \cite{Zanolli2006,Zanolli2006c}, photo-current (PC) \cite{Tragardh2006} or other electronic properties \cite{Bjork2005}.

The  situation is similar in the case of GaAs NW:s. When the (111)B GaAs surface is chosen as a substrate, the resulting NW:s consist of alternating wurtzite and zinc-blend segments \cite{Soshnikov2005}. Hence, both segments contribute to the optical and electronic properties of such a NW.

Theoretical studies based on total-energy calculations \cite{Crain1994} predict that the wurtzite polymorph ($wz, 2H$), having space group $P6_3mc$ ($C^6_{4v}$), is a metastable high-pressure modification of both the InAs and GaAs compounds. 
Hence, we are faced with new materials, InAs and GaAs in the wurtzite phase, for which no previous experimental study exists and only little theoretical work has been done. 

In this article we investigate the structural and electronic properties of these compounds by using  all-electron calculations based on Density-Functional Theory (DFT) \cite{Kohn1964_DFT}. Indeed, this  approach does not require a knowledge of material parameters and enables us to compute numerous properties ranging from the equilibrium lattice constant to the cohesive energy of the system. 

DFT in the Local Density Approximation (LDA) \cite{Kohn-Sham1965, Ulf1972} provides an accurate description of the ground state electronic structure of solids and such calculations are adequate tools for structural studies.  
The comparison of the calculated band structure to experimental excitation energies is, strictly speaking, not allowed but, nevertheless, provides important information on the electronic properties of the material. 
It is, for instance, well known that the experimental band gaps are usually underestimated by DFT. 
The excited-state properties of a many-electron system, such as the band structure, are better described within Many-Body Perturbation Theory (MBPT) as, for instance, by the GW approximation \cite{Hedin1965, Hedin1969} to the quasiparticle self-energy $\Sigma$. Still, a DFT calculation is a necessary step toward the GW calculation, since the LDA  eigenenergies and eigenfunctions are usually taken to be the zero:th order approximation  for the perturbative expansion. In particular, the LDA provides an accurate Hartree potential which always must be treated to infinite order in an extended system. Hence, the present band structure calculations allow us to investigate how well the LDA can work as a starting point for subsequent many-body calculations.

Although the LDA cannot describe the excited-state properties, it can nevertheless provide a good description of the properties of the occupied states \cite{Martin2004}. This means that we can extract useful informations from the valence part of the band structure as, for instance, the spin-orbit and - for the wurtzite compounds - the crystal field splittings.

Moreover, the all-electron band structures, calculated for the first time for the wurtzite phase, offer benchmarks for the corresponding pseudopotential calculations.

Related studies reported in the literature are performed within DFT, use pseudopotentials \cite{Wang2002, Wang2003}, and include the In 4{\it d} and the Ga 3{\it d} electrons among the core states. These approximations lead to an underestimation of the equilibrium lattice parameters and to a less accurate description of the band-structure. 
Here, we focus on a study of the ground state properties of InAs and GaAs in the wurtzite phase at the DFT/LDA level with a full potential description of the atomic species. 
For the calculations of quasiparticle band structures of these compounds we refer the reader to a subsequent paper \cite{Zanolli2006b}.




The article is organized as follows. After the description of the theoretical background underlying the calculations, we present the results
concerning the structure optimization of both the zinc-blende and the wurtzite phases of InAs. Then, we discuss the band structure obtained  in the non-relativistic and  in the scalar relativistic approximations, the latter with and without the inclusion of the spin-orbit (S-O) interaction. 
In the GaAs case, instead, we use the experimental lattice constants taken from TEM measurements on GaAs NW:s \cite{Jakob} to allow for a direct comparison with experiment. Indeed, since the deformation potential for the GaAs gap is very large \cite{Li2006}, small changes in the lattice constant have a strong influence on the band gap. Finally, we present the band structures for the GaAs compound calculated in the scalar relativistic approximation with and without the spin-orbit interaction.


\section{Methods of investigation}

The electronic structure calculations presented here are performed within DFT using the Full-Potential Linearized Augmented Plane Wave (FP-LAPW) method \cite{Slater1994}  as implemented in the WIEN2k software package \cite{WIEN2k}. 
For the exchange-correlation energy in the Local Spin Density Approximation (LSDA)  \cite{Ulf1972, Kohn1964_DFT, Kohn-Sham1965} we have chosen the electron gas data of Ceperley and Alder \cite{Ceperley1980} as parametrized by Perdew and Wang \cite{Perdew-Wang1992}.

The APW method \cite{Slater1937} is a procedure to solve the Kohn-Sham equation based on the division of the unit cell into two spatial regions: non-overlapping muffin-tin spheres centered at the atomic sites of radius $\rm R_{MT}$ and a remaining interstitial region. The basis set used consist of atomic partial waves and plane waves in the two regions, respectively.
Denoting  the cut-off wave number of the plane waves by $\rm K_{MAX}$, the number of basis functions is determined by the product $\rm R_{MT}\rm K_{MAX}$, where  $\rm R_{MT}$ is the smallest muffin-tin radius in the unit cell. The angular-momentum cut-off of the expansion of the wave functions in spherical harmonics within the atomic spheres is called $l_{MAX}$.

Starting from the original APW method, several improved versions of the method were developed as, for instance, the Linearized APW  method (LAPW) \cite{Andersen1975},  and, with the addition of local orbitals to the basis functions, the LAPW+LO \cite{DJSingh1991}, and the APW+lo \cite{Sjoststedt2000} methods.
Within the  WIEN2k code it is possible to use a mixed LAPW+LO and APW+lo basis \cite{Schwarz2002} to get the advantage of both schemes.
For the expansion inside the atomic spheres we have chosen the APW+lo basis for $l = 0, 1, 2$, the LAPW+LO for higher $l$:s and we have used  $l_{MAX} = 10$ as the highest $l$ value.


To ensure that no charge leaks outside the atomic spheres we have chosen $-9.0$ Ry for InAs and $-6.0$ Ry for GaAs
as the energy which separates the core and the valence states. In the InAs case, this choice guarantees that  the In $4s$, As $3d$ and all higher-energy states are treated as band states, {\it i.e.}, 36 valence electrons per InAs pair.
Instead, in GaAs, the valence band states start from the Ga $3d$ and the As $3d$, leading to 28 valence electrons per GaAs pair.

For both the InAs and the GaAs polymorphs we have performed convergence studies to determine the optimum value of $\rm R_{MT}\rm K_{MAX}$ and the number of $k$-points in the whole Brillouin zone (BZ). We have found that for $\rm R_{MT}\rm K_{MAX} = 11$ the total energy is converged to within less than 10 meV for both phases.
The number of $k$-points which ensures convergence is 1000 ($10\times10\times10$) in the {\it zb} and 405 ($9\times9\times5$) in the {\it wz} cases, respectively.
The chosen number of $k$-points corresponds to 73 $k$-points in the irreducible wedge of the Brillouin zone (IBZ)  for the {\it zb} structure and to 95 $k$-points for the {\it wz} structure. 
These $k$-meshes are used in a modified tetrahedron integration scheme \cite{Blochl1994} to determine which states are occupied. 
The Fourier expansion of the electron density in the interstitial region  ({\it i.e.} valence charge density) is cut-off at $\rm G_{MAX} = 14$, leading to 913 and 6524 plane waves for the $zb$ and the $wz$ structures, respectively. 


\section{Structure Optimization}
The parameters obtained in the convergence study were used to carry out the structure optimization of both InAs polymorphs in the way described below.

Due to the cubic symmetry, the equilibrium lattice constant $a_{zb}$ of the zinc-blende compound is determined by the minimization of the total energy of the system with respect to the volume of the unit cell. The energy vs volume curve 
is then fitted to the Murnaghan equation of state \cite{Murnaghan1944} giving the equilibrium parameters for the volume $V_0$ (and hence one lattice constant), the binding energy $E_0$ per pair of atoms, the bulk modulus $B$, and its pressure coefficient $B' = (dB/dp)_{p=0}$ . 

On the other hand, the wurtzite phase is characterized by two lattice parameters, namely $a_{wz}$ and $c$, and by the internal parameter $u$. Consequently,
the structure optimization of the wurtzite polymorph is carried out in four steps ordered according to the importance of the parameter under investigation. At first, the equilibrium volume is determined by assuming the ideal $c/a$ ratio. This gives a first estimate of the equilibrium $a_{wz}$ which is then used in the determination of the deviation of $c/a_{wz}$ 
from the ideal value. Then, the volume optimization is repeated using the new $c$ and $a_{wz}$  finally producing the equilibrium values. As the last step, the internal parameter $u$ is determined via force minimization.
We obtained $u/c = 0.3748$, which is very close to the ideal value (0.375).

\begin{figure}[t] 
\centering
\includegraphics[width=8cm]{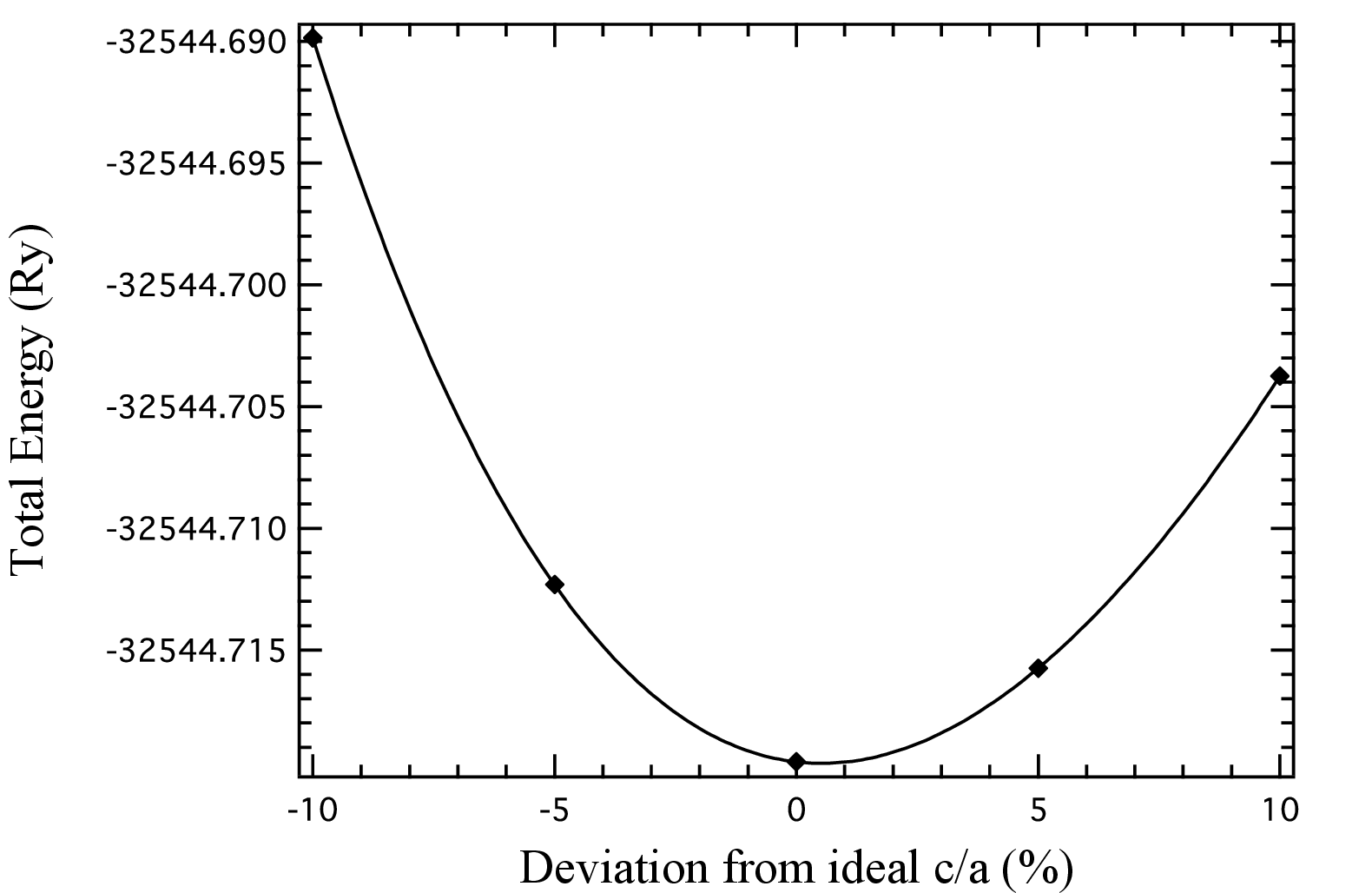}
\caption{Deviation from the ideal $c/a$ ratio for InAs in the wurtzite phase.}
\label{fig:ideal_c-a} 
\end{figure}

The deviation from the ideal $c/a_{wz}$ ratio is obtained by calculating the total energy of the {\it wz} polymorph for different values of the ratio and then fitting the so obtained curve to a $4^{th}$ order polynomial (Fig. \ref{fig:ideal_c-a}).   
In this way, we found $c/a = 1.64202$, which is slightly larger (by $0.5528~\%$) than the ideal value $\sqrt{8/3}$. 
The resulting equilibrium  value for  $c$ is 7.00118~{\AA}. 
The fact that the equilibrium value of $c/a$ is larger than the ideal one suggests  that
 the wurtzite phase is not the stable polymorph of InAs, in agreement with the rule stated in Ref.  \cite{Lawaetz1972}. 

\begin{table}[t]  
\caption{All-electron structural and energetic parameters for the two InAs polymorphs. The equilibrium volume and the binding energy refer to one InAs pair.}
\label{tab:Murnaghan_fit}
    \begin{center}
        \begin{tabular}{@{} |ccc|ccccccccc| @{}}
           \toprule
         &              &    &   &  $V_0 (a.u.^3)$  &  &  $B (GPa)$ &  &  $B'$    & &     $E_0$ (Ry)      &   \\ 
        	   \hline
         &  {\it wz} &    &      &    369.8984      &  &  59.6014    &  & 4.6170  & &  -16272.35981   &   \\ 
         &  {\it zb} &     &     &   372.2687       &  &  61.2977    &  & 4.1660   & &  -16272.35959   &  \\ 
           \hline \hline
       \end{tabular} 
    \end{center}
\end{table}


The coefficients of the Murnaghan fit of the energy vs volume curves are summarized in Table \ref{tab:Murnaghan_fit} for both polymorphs. By comparing the results obtained for two phases we find that the equilibrium volume per cation-anion pair of the wurtzite structure is smaller than that of the zinc-blende by 2.3703 a.u.$^3$, 
corresponding to a volume reduction of 0.637 \%.

\begin{figure}[t]   
\centering
\includegraphics[width=8cm]{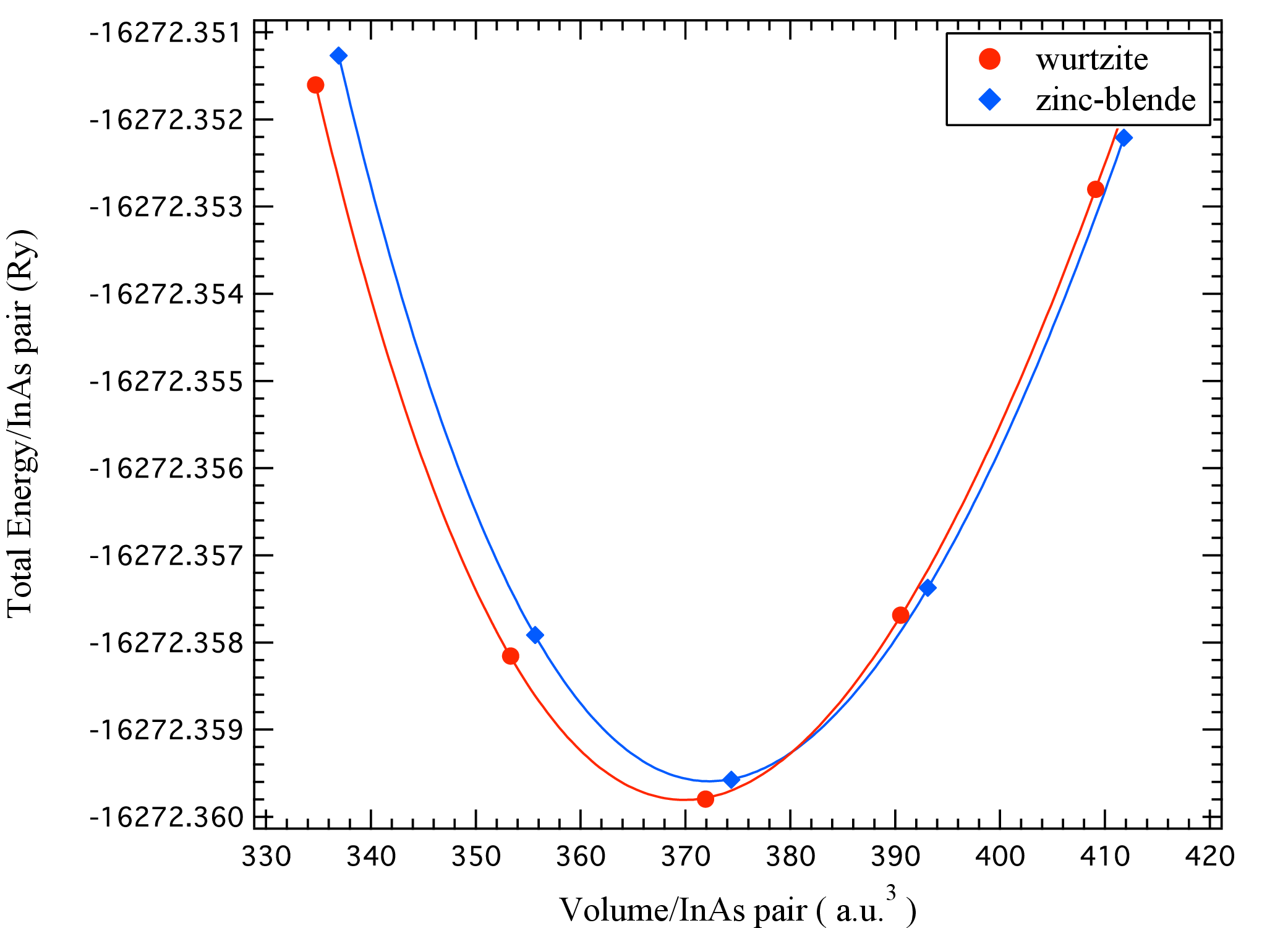}
\caption{Total energy versus volume normalized to one InAs pair for the zinc-blende (diamonds) and the wurtzite (circles) polymorphs. The continuous lines are the Murnaghan fit curves.}
\label{fig:Energy-Vol} 
\end{figure}

The result obtained for the binding energy is somewhat surprising. Contrary to the observed stability of the {\it zb} phase, we find that the binding energy of an InAs pair is larger in the {\it wz} than in the {\it zb} case  by 3~meV. This is clearly seen in the energy vs volume curves, normalized to one InAs pair, shown in Figure \ref{fig:Energy-Vol} for both phases.  We have checked this result by also calculating the total energy in the generalized gradient approximation (GGA) by Perdew-Burke-Ernzerhof (PBE) \cite{PBE}. In the GGA calculation we used the same lattice constant and convergence parameter as in the case of the LDA. The {\it wz} total energy was found to be 28~meV lower than that of the {\it zb}.
We note that the difference in the equilibrium total energies calculated within the LDA is actually smaller than the typical accuracy 
of calculations carried out within standard DFT codes, {\it i.e.} $\sim  30$~meV. We thus conclude that the relative stability of the two  phases cannot be determined within the present accuracy of the calculations.

As a further check, we have also performed the structure optimization by using Ultra Soft (US) pseudopotentials and plane waves \cite{Vanderbilt1985}, as implemented in the VASP code \cite{VASP}. The latter calculations are done by treating the In 4{\it d} states as either valence ({\it d}-val) or core ({\it d}-core) states. In the pseudopotential calculations we have used a Monkhorst-Pack \cite{Monk1976} grid centered on the $\Gamma$ point with a $11 \times 11 \times 11$  and $8  \times  8  \times  8$  mesh in reciprocal space for the {\it zb} and the {\it wz} phase, respectively. 
The convergence study gives the following kinetic energy cutoffs: 262~eV ({\it d}-val, for both polymorphs) and 222~eV and 202~eV ({\it d}-core) for the {\it zb} and {\it wz} cases, respectively. 
The  structure optimization was performed by minimization of the total energy.  In the wurtzite case we have assumed the ideal values for $c/a$ and $u/c$. 
We have thus found that the {\it zb} phase, having a binding energy lower than the {\it wz} by $\sim 18$~meV, is correctly predicted to be the stable phase of InAs. We also found that the equilibrium volume per InAs pair of the {\it wz} phase is smaller than the {\it zb} by 0.08\%. 

Clearly, the differing results is connected to the different method of calculation,  {\it i.e.} all-electron versus pseudopotential calculations. These tests thus provide an indication of the accuracy obtainable from calculations based on Ultra-Soft paseudopotentials. In our opinion, both sets of results must be considered correct in the sense that the total energy difference between them is of the same magnitude as the overall accuracy in the total energies from standard DFT codes.


\begin{table}[t]
\caption{Lattice constants for InAs in the zinc-blend and wurtzite phases. US denotes Ultra Soft pseudopotentials calculations with the In 4{\it d} electrons included among the valence ({\it d}-val) or core ({\it d}-core) states. The $c/a$ and $u/c$ values were assumed to be the ideal ones ({\it i.e.} 1.633 and 0.375, respectively) in pseudopotential calculations. The percentage deviation of $c/a$  from the ideal value is 0.5528\% in the full potential calculation.}
\label{tab:latt_const}
    \begin{center}
       \begin{tabular}{|ccc|ccc|ccc|ccc|ccc|ccc|}
           \hline \hline
       &                        &     && $a_{zb}$ ({\AA}) &   && $a_{wz}$ ({\AA}) &   &&  $c$ ({\AA}) &    &&  $c/a$ &     	 && $u/c$ &\\ 
        	   \hline
       & exp.                 &     &&   6.0542              &   && 4.2839  &    		 && 6.9954 	&	&& 1.633	&	 &&  0.3750 &\\         
       & all-electr.        &     &&   6.0428              &   && 4.2638  &  		&& 7.0018   	&	&& 1.642	&	 && 0.3748 &\\
       & US {\it d}-val   &     &&   6.0329              &   && 4.2663  &  		&& 6.9669    	&	&&  ideal 	&	 &&  ideal &  \\
       & US {\it d}-core &     &&   5.8023              &   && 4.1060  & 		 && 6.7051	& 	&&  ideal 	&	 &&  ideal & \\
           \hline \hline 
       \end{tabular} 
    \end{center}
\end{table}

The calculated equilibrium lattice constants (in both the full and pseudo potential method) are 
displayed in Table \ref{tab:latt_const} together with the experimental values. 
The experimental values for the wurtzite lattice constants are taken from TEM measurements \cite{Zanolli2006c}. For the zinc-blende structure the experimental parameter measured at room temperature \cite{Landolt1982} has been extrapolated to zero degrees Kelvin. 
We thus find that the all-electron calculations and the pseudopotential calculations with the In 4d treated as valence electrons give very similar results and they slightly underestimate ($\sim 0.01 - 0.02$ {\AA}) the experimental values for the lattice parameters, as is usually the case within the LDA. 
The results of the pseudopotential calculation with the In 4{\it d} electrons frozen into the core are, instead, rather far from the experimental values. This shows the importance of treating the {\it d}-electrons as valence states in order to obtain a realistic description of the InAs compound. 
In the {\it wz} case the results reported in the literature \cite{Wang2002} are calculated using HGH pseudopotentials \cite{HGHpsp} constructed by treating the {\it d} electrons as core states. The results are:  $a_{wz} = 4.192$~\AA, $u = 0.3755$, $c = 0.6844$~\AA. 
Also results for the {\it zb} case are reported in the same article using the same approach, leading to $a_{zb} = 5.921$~{\AA}. 
Other calculations based on the all-electron approach have given the results $a_{zb} = 6.063$~{\AA} \cite{Massidda1990} and $a_{zb} = 6.084$~{\AA} \cite{Srivastava1988} in the case of  InAs in the {\it zb} phase.


The last step in the structural optimization is the search for that value of the internal parameter $u$ for which the force on the nuclei vanishes. We searched for this point starting from the {\it wz} structure with the equilibrium $c$ and $a_{wz}$ lattice constants. The search was carried out by means of a minimization routine from the PORT library \cite{PORT} 
choosing 0.5 mRy/Bohr as a tolerance on the force. We obtained $u/c = 0.37481$, {\it i.e.} 0.05\% smaller than the ideal value. 

\section{Band structures}

All the LDA band structures presented in this article are calculated by solving the Kohn-Sham equation \cite{Kohn-Sham1965} at the experimental lattice parameters, along lines in $k$-space which include high symmetry points. We have performed such calculations for both the {\it wz} and the {\it zb} polymorphs of InAs and GaAs.

The motivations for calculating the full-potential LDA band structures are as follows. 
The all-electron bands
(i)  allow us to investigate how well the LDA can function as a zero:th order approximation for the perturbation expansions of MBPT;
(ii) allow us to assess the effects of different relativistic approximations;
(iii) provide a benchmark for pseudopotential calculations based on LDA;
(iv) provide a good approximation to the electronic structure of the occupied states ({\it i.e.} valence band states), thus allowing us to estimate the spin-orbit and the crystal field splittings for the new wurtzite structures.

The WIEN2k code allows for the inclusion of relativistic effects at different levels of approximation: non-relativistic, scalar relativistic without spin-orbit interaction and approximate fully-relativistic with the inclusion of the spin-orbit interaction. 
The core states are treated fully relativistically \cite{Desclaux}, unless otherwise specified.
In the case of the valence states the relativistic effects are included via a scalar relativistic approximation \cite{Koelling1977} whereafter the spin-orbit interaction can be added 
by using the scalar relativistic eigenfunctions as a basis \cite{Macdonald1980, Singh1994}. 
In order to overcome the problem associated with the averaging procedure inherent in the scalar relativistic approximation additional local orbitals of definite $j$ character have been added to the basis set \cite{Kunes2001}. 


\begin{figure*}[t]
\centering
\includegraphics[width=18cm]{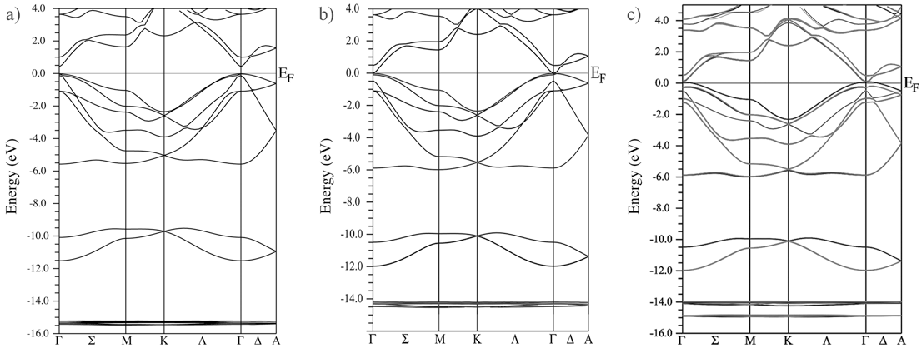}
\caption{Full potential band structures of InAs in the wurtzite phase in different relativistic approximations: (a) non-relativistic, (b) scalar-relativistic, and (c) scalar-relativistic with the inclusion of spin-orbit coupling.}
\label{fig:InAs_wz_bands} 
\end{figure*} 

When the  spin-orbit coupling is taken into account, the corresponding Hamiltonian is diagonalized using scalar relativistic eigenfunctions in a range of 20 Ry centered around the top of the valence band. This is necessary because the S-O splitting in InAs  has a strength comparable to the energy gap 
and can, thus, not be considered as a ``small'' correction to the scalar relativistic bands.
In the case of GaAs the S-O spitting is smaller than the energy gap (at low temperature, for the {\it zb} phase: 0.33~eV and 1.519~eV, respectively  \cite{Landolt1982}), but still the effect is important and so we use the same approximation as for the InAs.

\subsection{InAs band structures}

At first, the band structures of InAs in both the {\it zb} and the {\it wz} phases were calculated within the non-relativistic approximation. We then found that InAs is correctly predicted to be a semiconductor with energy gaps of 0.395~eV and 0.444~eV in the {\it zb} and the {\it wz} cases, respectively. 
We are grateful to Prof. Almbladh for corroborating these results using his own  LAPW code \cite{COA}. Hence, within the LDA, the energy gap of the {\it wz} polymorph is 49~meV larger that that of the {\it zb}. In the {\it zb} case, the three $p$-like states constituting the valence band maximum (VBM) are degenerate at the $\Gamma$ point. In the {\it wz} case two of these states show a crystal field splitting of 91~meV. The resulting band structure for the {\it wz} polymorph is displayed in Fig.~\ref{fig:InAs_wz_bands}a.

%


The scalar relativistic band structures of the two polymorphs were then calculated without including the spin-orbit interaction. Within this approach the InAs energy gap  is not present anymore. Instead, we found a ``negative gap" or ``wrong band ordering"  for both polymorphs. Indeed, in the {\it zb} case, at the $\Gamma$ point, the three $p$-like degenerate states at the top of the valence band have a higher energy than the $s$-like conduction band minimum (CBM), resulting in an energy gap of $\Gamma_{1c} -  \Gamma_{15v} = -0.467$~eV. The {\it wz} phase exhibit the same problem and the calculated energy gap is $\Gamma_{1c} -  \Gamma_{1v} = - 0.359$~eV. The crystal field splitting can be identified as a splitting of $84$~meV between the the $\Gamma_{6v}$ and  $\Gamma_{1v}$ valence band levels.
The scalar relativistic band structure for InAs in the {\it wz} phase is shown in Fig. \ref{fig:InAs_wz_bands}b.



The main reason for this un-physical result has to be ascribed to the notorious limitations of the LDA in reproducing band structures. Since the calculated energy gaps are typically underestimated, this may result in the ``negative gap" problem when the semiconductor under investigation has a small energy gap. This is indeed the case in InAs, for which the experimental energy gap of the {\it zb} polymorph at zero Kelvin is 0.415~eV \cite{Landolt1982}.

We should also consider the fact that the binding energy of the In 4{\it d} states is under-estimated by the LDA  \cite{Zakharov1994}. As a consequence the {\it d} states are too close to the VBM, resulting in a too large repulsion between the {\it p} and the {\it d} states. This has the effect of shifting the VBM toward higher energies, hence contributing to the reduction of the energy gap. 


Another source of error can be found in the scalar relativistic treatment of InAs. Indeed, in this semiconductor, the relativistic effects are of the same order of magnitude as the energy gap, as can be seen by considering the fact that the spin-orbit splitting at zero Kelvin is 0.38 eV \cite{Landolt1982}. 

We have, therefore, improved our relativistic description of the system by including the spin-orbit interaction as a correction to the scalar relativistic band structure. 
This approach led to spin-orbit splittings in the {\it zb} and the {\it wz} cases of 0.355~eV and 0.336~eV, respectively. In addition, we obtained a crystal field splitting of 48~meV in the {\it wz} polymorph.
Unfortunately, the inclusion of spin-orbit interaction did not solve the problem with the  ``wrong band ordering". We obtained the  ``negative energy gaps'' $-0.453$~eV and $-0.506$~eV for the {\it zb} and the {\it wz} phases, respectively. The band structure with the inclusion of the spin-orbit interaction is shown in Fig. \ref{fig:InAs_wz_bands}c  for the {\it wz} polymorph. The results for the InAs case are collected in Tab. \ref{tab:InAs}.


\begin{table}[hbt]
\caption{The experimental and calculated energy band gaps, spin-orbit splittings and crystal field splittings (only for {\it wz}) for the two polymorphs of the InAs. The tabulated crystal field splitting is calculated with the inclusion of the S-O interaction. All the energies are in eV and the experimental values are from Ref. \cite{Landolt1982}.}
\label{tab:InAs}
    \begin{center}
       \begin{tabular}{|c|c|ccc|ccc|ccc|c|}
           \hline \hline
         	         &  $E_{gap}$  &&  $E_{gap}$   && & $E_{gap}$  &   &&   S-O   &&  Crystal  \\ 
         	         &  non-rel       &&   scal. rel        && & scal rel      &   &&  splitt.  &&  field  \\ 
         	         &                    &&   w.out S-O   && & with S-O     &   &&           &&  splitt.  \\ 
        	   \hline
           zb          & 0.395         &&        -0.467     && &    -0.453       &  &&  0.336  &&     none  \\         
          zb (exp) &                  &&       0.542    && & 0.415         &  &&  0.38    &&    none   \\
          wz         & 0.444        &&    -0.359       && &    -0.506      &  &&   0.355 &&  48     \\
           \hline \hline
       \end{tabular} 
    \end{center}
\end{table}

\subsection{GaAs band structures}

The band structure calculations for GaAs were performed by using the values of lattice constants measured via TEM \cite{Jakob} on single NW:s having a diameter of 110~nm. We used the lattice constants $a_{zb} = 5.653$~\AA, $a_{wz} = 3.994$~\AA, $c = 6.528$~{\AA} and $u/c = 3/8$, {\it i.e.} the ideal $u/c$ value.
As mentioned in the introduction, since the deformation potential of GaAs is very strong the calculated band gap depends in a critical way on the lattice constant. Hence, in order to allow a better comparison with experiment, we choose to work with the experimental lattice parameters. 

We tested this choice by calculating the relativistic band structure (without taking into account the spin-orbit interaction) of GaAs in the {\it zb} phase with both the TEM lattice constant and the equilibrium lattice constant $a_{zb, eq} = 5.6$~{\AA} taken from Ref. \cite{Arabi2006}.
The resulting band gaps are 280~meV and 493~meV, respectively. Hence, a decrease in the lattice constant of 0.94\% results in an increase of the band gap of  43.2\%.

\begin{figure}[t]
\centering
\includegraphics[width=6cm]{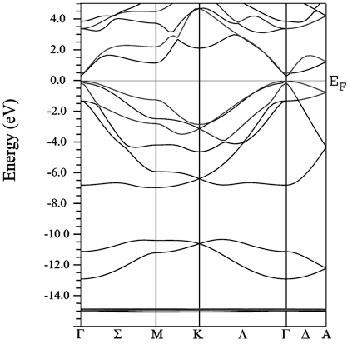}
\caption{Scalar-relativistic band structure of GaAs in the wurtzite phase.}
\label{fig:GaAs_wz} 
\end{figure}

\begin{figure}[htb]
\centering
\includegraphics[width=6cm]{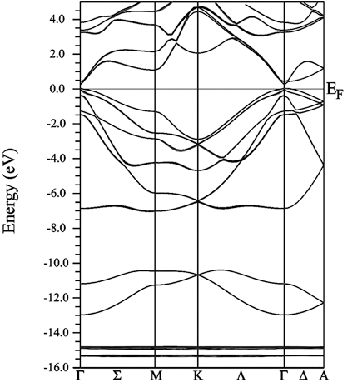}
\caption{Scalar-relativistic band structure of GaAs in the wurtzite phase with the inclusion of spin-orbit coupling.}
\label{fig:GaAs_wz_SO} 
\end{figure}

The scalar relativistic band structures of both the {\it zb} and the {\it wz} phases of GaAs were calculated using the same procedure as in the case of InAs. The main difference with respect to InAs is that GaAs is correctly predicted to be a semiconductor, even though the LDA gap is very small compared to experiment. This behavior can be explained with arguments analogous to those used in the discussion of the InAs band structures. 
The band structures calculated for the {\it wz} polymorph without and with the inclusion of the spin-orbit interaction are shown in Figs. \ref{fig:GaAs_wz} and \ref{fig:GaAs_wz_SO}, respectively.

\begin{table}[hbt]
\caption{The experimental and calculated energy band gaps, spin-orbit splittings and crystal field splittings (only for {\it wz}) for the two polymorphs of the GaAs. The tabulated crystal field splitting is calculated with the inclusion of the S-O interaction. All the energies are in eV and the experimental values are from Ref. \cite{Landolt1982}.}
\label{tab:GaAs}
    \begin{center}
       \begin{tabular}{|ccc|ccc|ccc|ccc|c|}
           \hline \hline
       &  	         &        &&  $E_{gap}$   && & $E_{gap}$  &   &&   S-O   &&  Cryst. field  \\ 
       &  	         &        &&   w.out S-O   && & with S-O     &   &&  splitt.  &&  splitting  \\ 
        	   \hline
       &   zb          &       &&  0.280           && & 0.167         &  && 0.339    &&     none  \\         
       &   zb (exp) &       &&  1.629          && & 1.519          &  && 0.33      &&    none   \\
       &    wz         &       &&  0.327          && & 0.211          &  && 0.340   &&      79  \\
           \hline \hline
       \end{tabular} 
    \end{center}
\end{table}

Without the S-O interaction we find that the band structure of the {\it wz} phase is characterized by a band gap of 327~meV and a crystal field splitting of 131~meV. 
When the S-O interaction is taken into account, the band gap is decreased to 167~meV for the {\it zb} and to 211~meV for the {\it wz} phase. The calculated spin-orbit splitting is 339~meV  and 340~meV in the {\it zb} and the {\it wz} cases, respectively, and the crystal field splitting in the {\it wz} band structure is 79~meV. 
All these results are collected in Tab. \ref{tab:GaAs}.

\section{SUMMARY AND CONCLUSIONS}

We have reported full-potential all-electron studies of InAs and GaAs in both the wurtzite and the zinc-blende phases. 
The structural optimization of both polymorphs of InAs is carried out and the result is in good agreement with the TEM measured lattice parameters. We have found that, within the present accuracy of standard density functional calculations, it is not possible to decide which of the phases of InAs is the stable one. Our total energy difference between the phases is  $\sim 3$~meV and much smaller than the numerical accuracy of the code ($\sim 30$~meV).

In the case of GaAs, the band structures were calculated at the experimental lattice constants measured via TEM on NW:s having partially wurtzite structure.

The LDA band structures have been computed for both polymorphs of InAs and GaAs using different relativistic approximations. At first, results from the non-relativistic approach have been presented for InAs. Then we performed calculations using an approximate relativistic model in which the core states are treated fully relativistically while the valence states are described by a scalar relativistic model. Spin-orbit interaction was then added as a correction. 
When the bands are calculated relativistically we find that InAs is wrongly predicted to be a zero-gap semiconductor while InAs is a finite gap semiconductor within the non-relativistic approximation. Instead GaAs is correctly predicted to be a semiconductor in the scalar relativistic approximation, even tough the calculated band gaps are considerably smaller than the experimental one.

Spin-orbit splittings and crystal field splittings for the new polymorph of InAs and GaAs have been calculated for the first time.



\begin{acknowledgments}
We thank the Nanoquanta network of Excellence (contract number NMP4-CT-2004-500198) for support. We also thank the  Photon-Mediated Phenomena (PMP) Research Training Network (contract number HPRN-CT-2002-00298) for supporting this research.
Finally, we thank Prof. C.-O. Almbladh for assistance and helpful comments during the course of this work and Dr. Jakob B. Wagner for TEM imaging.
\end{acknowledgments}

%


\end{document}